\documentclass[twocolumn]{revtex4}
\usepackage{amsmath,amssymb,graphicx,color}
\usepackage[mathscr]{euscript}

\def\babc{\begin{subequations}}
\def\eabc{\end{subequations}}
\def\be{\begin{equation}}
\def\ee{\end{equation}}
\def\ba{\begin{array}}
\def\ea{\end{array}}
\def\nn{\nonumber}

\def\bd{\boldsymbol{\delta}}
\def\br{\boldsymbol{r}}
\def\bk{\boldsymbol{k}}
\def\boa{\boldsymbol{a}}
\def\dk{d{\bk}}

\begin{document}

\title{The quantum group, Harper equation and the structure of Bloch eigenstates\\
on a honeycomb lattice}

\author{Merab Eliashvili$^{1,2}$, George I. Japaridze$^{3}$ and George Tsitsishvili$^{1,2}$}
\affiliation{
$^{1}$Faculty of Exact and Natural Sciences, Tbilisi State University, Chavchavadze Ave. 3, Tbilisi 0128, Georgia\\
$^{2}$Razmadze Mathematical Institute, Tbilisi State University, Chavchavadze Ave. 3, Tbilisi 0128, Georgia\\
$^{3}$College of Engineering, Ilia State University, Cholokashvili Ave. 3-5, Tbilisi 0162, Georgia}

\begin{abstract}
The tight-binding model of quantum particles on a honeycomb lattice is investigated in the presence of homogeneous magnetic field.
Provided the magnetic flux per unit hexagon is rational of the elementary flux, the one-particle Hamiltonian is expressed in terms
of the generators of the quantum group $U_q(sl_2)$. Employing the functional representation of the quantum group $U_q(sl_2)$
the Harper equation is rewritten as a systems of two coupled functional equations in the complex plane. For the special values of
quasi-momentum the entangled system admits solutions in terms of polynomials. The system is shown to exhibit certain symmetry
allowing to resolve the entanglement, and basic single equation determining the eigenvalues and eigenstates (polynomials) is obtained.
Equations specifying locations of the roots of polynomials in the complex plane are found. Employing numerical analysis the roots of
polynomials corresponding to different eigenstates are solved out and the diagrams exhibiting the ordered structure of one-particle
eigenstates are depicted.
\end{abstract}

\maketitle

\section{Introduction}

The problem of quantum particles in the presence of a periodic potential and a uniform magnetic field has been the subject of intensive
studies for decades. Azbel \cite{azbel} was the first who pointed out that the spectral properties of two-dimensional lattice particles
have a sensitive dependence on the flux through the plaquette. This observation has been exploited later by Hofstadter \cite{hofstadter}
who found the exotic structure of the one-particle energy spectrum of planar particles on square lattice in magnetic field. The same study
was extended later for triangular lattice \cite{claro79}, generalized square lattices \cite{claro81,hasegawa} and for the honeycomb lattice
\cite{semenoff,rammal}. These studies firmly established the fractal structure of the aforementioned energy spectrum, whose rich and
complex nature originates from the presence of two, not necessarily commensurate periods. The first is given by the lattice structure and
the second is determined by the magnetic field. The relevant parameter which determines the band structure is the ratio $\Phi/\Phi_0$
where $\Phi$ is the magnetic flux per elementary plaquette, and $\Phi_0=2\pi(\hbar/e)$ is the magnetic flux quantum.

In 1994 Wiegmann and Zabrodin pointed out \cite{wiegmann} that the Hamiltonian responsible for the original result of Hofstadter is closely
related to the quantum group $U_q (sl_2)$ (for mathematical treatment see \cite{faddeev}). Namely, it was shown, that the Hamiltonian is
expressible in terms of $X^\pm$ generators of quantum group $U_q(sl_2)$: $\mathscr H=X^++X^-$ with a deformation parameter $q$
determined by the applied magnetic field. Employing the functional representation of $U_q(sl_2)$ in the space of polynomials, the Harper equation
was reformulated into the functional form where the one-particle eigenstates appeared as polynomials. The zeros of polynomials unambiguously
specifying the one-particle wave functions were shown to be determined by the Bethe ansatz equations \cite{wiegmann}. The fact that eigenstates
are related to certain polynomials associated with the quantum group $U_q (sl_2)$ and with the Bethe ansatz equations provides with the possibility
to systematically study the structure of eigenstates. It should be noted that so far the studies were mainly concentrated on the structure of spectrum.

Hatsugai {\it et al.} \cite{hatsugai} investigated the Bethe ansatz equations derived in \cite{wiegmann} and found out that the zeros of those polynomial
are not scattered randomly, but are located along concentric circles on a complex plane, thereby exhibiting nontrivial and ordered structure of eigenstates.

Alongside with the eigenvalue problem on a square lattice, the analogous problem has been considered for a honeycomb lattice as well \cite{semenoff,rammal}.
The common observation is that the energy spectrum differs from the one of square lattice and is equally highly nontrivial. After the connection between
the quantum group $U_q(sl_2)$ and the problem of square lattice particles in magnetic field was found out in \cite{wiegmann}, the similar approach has been
applied to honeycomb lattice by Kohmoto and Sedrakyan \cite{kohmoto}.

In contrast to the square lattice, the honeycomb is not a Bravais lattice, but consists of two interpenetrating triangular lattices ("$A$" and "$B$" sublattices)
with one lattice point of each type per unit cell. Therefore the one-particle Hamiltonian possesses additional $(2\times2)$-matrix structure. In  terms of the
generators of $U_q(sl_2)$ it appears as
\be
\mathscr H=\left(\ba{cc} 0 & \mathbb I+X^- \\\\ \mathbb I+X^+ & 0\ea\right)
\ee
and carries the anti-diagonal structure with respect to $(2\times2)$-matrix indices. This causes certain inconvenience (eigenvalue problem consists of two
entangled equations) which can be overcome by considering $\mathscr H^2$ instead of $\mathscr H$ since the former is diagonal in $(2\times2)$-matrix
indices. Taking advantage of this observation, the authors of \cite{kohmoto} obtained the analogue of the Bethe ansatz equation derived for square lattice
in \cite{wiegmann}. The Hamiltonian (1) describes particle hoppings from a site to the three nearest neighboring ones. The squared operator $\mathscr H^2$
contains second-order terms describing double-hoppings $A\to B\to A$ and $B\to A\to B$. As a result, the Bethe-like equations derived in \cite{kohmoto}
look quite cumbersome, and the nesting procedure has been applied with the aim to reduce them to the set of simpler equations.

In this paper we develop the distinct approach to the eigenvalue equation for the Hamiltonian (1) by revealing its novel symmetry. This symmetry allows
to carry out certain analytic calculations right for $\mathscr H$ with no necessity of involving $\mathscr H^2$. Based on this symmetry we reduce the
system of entangled equations to a single equation thus avoiding the necessity of involving $\mathscr H^2$. The obtained single equation corresponds to
the elementary hoppings $A\rightleftarrows B$ and appears perceptibly simpler than the one derived in \cite{kohmoto} from $\mathscr H^2$.

The paper is organized as follows. In the forthcoming section we comment on the derivation of one-particle Hamiltonian from the tight-binding model on
honeycomb lattice in magnetic field. In Section III we rewrite this Hamiltonian in terms of the quantum group $U_q(sl_2)$ and discuss the representations
of $U_q(sl_2)$. Section IV deals with Harper equation in polynomial representation. Since the honeycomb consists of two triangle Bravais sublattices,
the Harper equation appears in the form of the system of two entangled equations. This system is shown to possess certain type of symmetry which permits
to decouple them and to obtain a single equation describing the elementary process of nearest neighbouring hopping. Solutions to this equation are polynomials
with real coefficients and Bethe-like equations determining the zeros of polynomials are derived. In Section V the results of numeric analysis are presented
and the locations of zeros in the complex plane are depicted exhibiting nontrivial and ordered structure of eigenstates.

\section{Quantum particle on a honeycomb lattice in magnetic field}

Honeycomb lattice is built up of two-dimensional array of hexagonal unit cells of the side $a/\sqrt3$, with atoms at the vertices. Such a structure is encountered
in solid state physics in various crystals, while the ideal realization of the two-dimensional honeycomb lattice is graphene \cite{geim}. The unit cell is a rhombus of
the side $a$ with angles $\pi/3$ and $2\pi/3$ at its vertices. Each unit cell consists of two atoms, $A$ and $B$ as indicated in figure 1.

We study the tight-binding model on a honeycomb lattice with the nearest neighbouring hoppings only. In the presence of a homogeneous magnetic field
the Hamiltonian under consideration is given by
\be
H=\sum_{n,\br}\Big[e^{-i\gamma_n(\br)}c^\dag_{B}(\br+\bd_n)c^{}_{A}(\br)\Big]+h.c.,
\ee
where the sum with respect to $\br$ is implied over the sites $\br=j_1\boa_1+j_2\boa_2$. Here $c^\dag_{A}(\br)$ ($c^{}_{A}(\br)$)
and $c^\dag_{B}(\br+\bd_n)$ ($c^{}_{B}(\br+\bd_n)$) are the particle creation (annihilation) operators on site $\br$ of the sublattice
"$A$" and on site $\br+\bd_n$ of the sublattice "$B$", respectively.

Magnetic field $\mathcal B$ is included in the Hamiltonian via the Peierls phases
\be
\gamma_n(\br)=\frac{e}{\hbar}\int_{\br}^{\br+\bd_n}\boldsymbol{\mathcal A}d\boldsymbol{l},
\ee
where the vector-potential is taken in the Landau gauge $\boldsymbol{\mathcal A}=(-\mathcal By,0)$.

\begin{widetext}

\begin{figure}[h]
\includegraphics{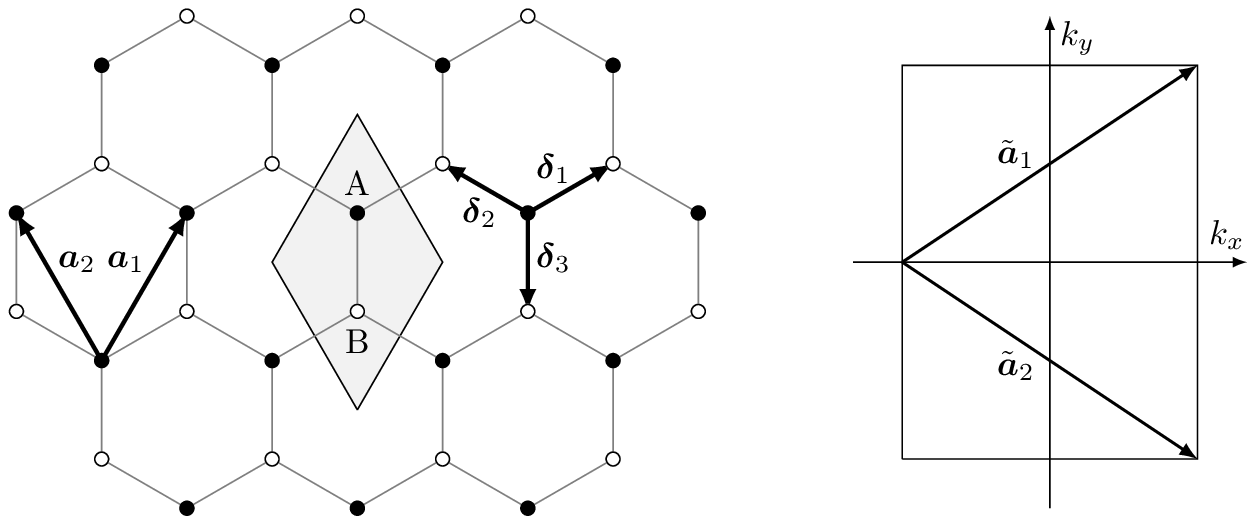}
\caption{Honeycomb lattice (left) consists of two Bravais sublattices $A$ and $B$. The $A$-sites are located at
$\br=j_1\boa_1+j_2\boa_2$ where $\boa_{1,2}=\frac12(\pm1,\sqrt3)a$, and $j_{1,2}$ are integers. The three
$B$-sites, nearest to a given $A$-site are located at $\br+\bd_{1,2,3}$. Brillouin zone (right) is arranged in the
rectangular form. The vectors $\tilde{\boa}_{1,2}$ set by $\boa_i\tilde{\boa}_j=2\pi\delta_{ij}$ determine the
structure of the Brillouin zone.}
\end{figure}

Rewriting (2) in the Fourier representation we find
\begin{align}
H=\int\Big[c^\dag_{B}(\bk)c^{}_{A}(\bk)
+e^{-i\bk\boa_1}c^\dag_{B}(\bk)c^{}_{A}(\bk+\bk_0)
+e^{-i\bk\boa_2}c^\dag_{B}(\bk+\bk_0)c^{}_{A}(\bk)\Big]\dk+h.c.,
\end{align}
\end{widetext}
where $c_{A}(\bk)$ and $c_{B}(\bk)$ are the Fourier transforms of $c^{}_{A}(\br)$ and $c^{}_{B}(\br+\bd_n)$ respectively,
and the integration covers the first Brillouin zone ({\tt FBZ}).

The vector $\bk_0$ is related to the magnetic field
\be
\bk_0=\frac{2\pi}{\sqrt3a}\frac{\Phi}{\Phi_0}(0,1).
\ee
where $\Phi$ and $\Phi_0$ are the magnetic flux per elementary hexagon and the magnetic flux quantum, respectively.

We consider
\be
\frac{\Phi}{\Phi_0}=\frac{\nu}{N},
\ee
where $\nu$ and $N$ are coprime integers, and concentrate on the odd values of $N$.

Then the Hamiltonian (4) can be presented as follows (see Appendix A)
\babc
\begin{align}
H&=\int_{\tt MBZ}\Psi^\dag(\bk)\mathscr H(\bk)\Psi(\bk)\dk,\\
\nn\\
\mathscr H(\bk)&=\left\lgroup\ba{cc}0&\mathbb I+X^-(\bk)\\\\\mathbb I+X^+(\bk)&0\ea\right\rgroup,
\end{align}
\eabc
where $\Psi(\bk)$ is a $(2N)$-component column, and the integration covers the magnetic Brillouin zone,
which is the $N$'th part of the first Brillouin zone.

The $N\times N$ matrices $X^\pm(\bk)$ are given by
\babc
\begin{align}
X^+(\bk)&=e^{-i\bk\boa_1}\beta^\dag Q+e^{-i\bk\boa_2}Q\beta,\\
\nn\\
X^-(\bk)&=e^{+i\bk\boa_1}Q^\dag\beta+e^{+i\bk\boa_2}\beta^\dag Q^\dag,
\end{align}
\eabc
where
\be
\beta=\left\lgroup\ba{cccccc}
0 & 1 & 0 & \cdots & 0 & 0\\
0 & 0 & 1 & \cdots & 0 & 0\\
0 & 0 & 0 & \cdots & 0 & 0 \vspace*{-1mm}\\
\vdots & \vdots & \vdots &  & \vdots & \vdots\\
\vspace*{-3.5mm}\\
0 & 0 & 0 & \cdots & 0 & 1\\
1 & 0 & 0 & \cdots & 0 & 0
\ea\right\rgroup
\ee
and
\be
Q={\tt diag}\big(q^1,q^2,\ldots,q^N\big)
\ee
with
\be
q=e^{+i\pi(\nu/N)}.
\ee

\section{The quantum group $U_q(sl_2)$}

Using the relation $Q^\dag\beta Q=q^2Q\beta Q^\dag$ we find
\babc
\begin{align}
\big[X^+,X^-\big]&=i^2(q-q^{-1})(K-K^{-1}),\\
\nn\\
KX^\pm K^{-1}&=q^{\pm2}X^\pm,
\end{align}
\eabc
where
\babc
\begin{align}
K(\bk)&=qe^{+i\bk(\boa_1-\boa_2)}Q\beta Q^\dag\beta,\\
\nn\\
K^{-1}(\bk)&=q^{-1}e^{-i\bk(\boa_1-\boa_2)}\beta^\dag Q\beta^\dag Q^\dag.
\end{align}
\eabc

The relations (12) constitute the definition of the quantum group $U_q(sl_2)$ which is the deformation of usual $sl_2$ (see \cite{kassel}) with
$q$ referred to as the deformation parameter (we adopt the normalization of $X^\pm$ which slightly differs from the standard one). Note that
particular values of $\bk$ and $\boa_{1,2}$ do not affect the structure (12).

\subsection{Cyclic and highest weight representations}

For every complex $q\ne\pm1$ the quantum group $U_q(sl_2)$ possesses the highest weight representations analogous to those of ordinary $sl_2$.
Furthermore, when $q$ is the root of unity, the so called cyclic representation also does exist.

Due to $q^N=e^{i\pi\nu}=(-1)^\nu$ the cases when $\nu$ is either even or odd slightly differ from each other. Hereafter we discuss the even values
of $\nu$. The case of odd values leads to identical outcome, and the corresponding technical details are collected in Appendix B.

Introduce the states $\psi_1,\psi_2,\ldots,\psi_N$ in the form of the $N$-component columns
\be
\psi_j=N^{-\frac12}\big\{q^j,q^{2j},q^{3j},\ldots,q^{Nj}\big\}{}^{\tt T}.
\ee
These states form the orthonormal complete set
\babc
\begin{align}
\sum_{n=1}^N(\psi^\dag_i)_n(\psi^{}_j)_n&=\delta_{ij},\\
\nn\\
\sum_{j=1}^N(\psi^\dag_j)_m(\psi^{}_j)_n&=\delta_{mn},
\end{align}
\eabc
and can be used as a basis in the space of representation.

The action of matrices $\beta$ and $Q$ on the states (14) is expressed by the following relations
\babc
\begin{align}
\beta\psi_j&=q^j\psi_j,\\
\nn\\
\beta^\dag\psi_j&=q^{-j}\psi_j,\\
\nn\\
Q\psi_j&=\psi_{j+1},\\
\nn\\
Q^\dag\psi_j&=\psi_{j-1},
\end{align}
\eabc
where the cyclic identification $\psi_{N+j}=\psi_j$ is understood in (16c) and (16d). This is possible due to $q^N=1$.

Using (16) we find
\babc
\begin{align}
X^+\psi_j&=q^{-\frac12}e^{-\frac{i}{2}\sqrt3k_ya}t_{j+1}\psi_{j+1},\\
\nn\\
X^-\psi_j&=q^{+\frac12}e^{+\frac{i}{2}\sqrt3k_ya}t_j\psi_{j-1},\\
\nn\\
K\psi_j&=e^{+ik_xa}q^{+2j}\psi_j\\
\nn\\
K^{-1}\psi_j&=e^{-ik_xa}q^{-2j}\psi_j,
\end{align}
\eabc
where
\be
t_j=e^{+\frac{i}{2}k_xa}q^{j-\frac12}+e^{-\frac{i}{2}k_xa}q^{-j+\frac12}.
\ee

Due to (17a) and (17b) the operators $X^+$ and $X^-$ can be regarded as rising and lowering ones respectively.

From (17a) and (17b) we find
\be
{\tt Det}X^\pm=q^{\mp\frac12N}e^{\mp\frac{i}{2}N\sqrt3k_ya}(t_1t_2\cdots t_N).
\ee

For those values of $k_x$ when $t_j\ne0$ for any $j$, the operators $X^\pm$ possess only nonvanishing eigenvalues.
In other words, there is neither highest nor lowest weight states. Hence (8) and (13) form the cyclic representation of (12).
Acting on $\psi_j$ by $X^+$ we obtain $\psi_{j+1}$, and so on until we arrive to $\psi_N$. The subsequent action takes
us back to $\psi_1$. The similar is true for $X^-$.

On the other hand, for the value of $k_x$ set by $e^{+\frac{i}{2}k_xa}=\pm i q^{\frac12-j_0}$ we have $t_{j_0}=0$ and
the cyclic representation turns into the highest weight one, {\it e.g.} setting $e^{\frac{i}{2}k_xa}=iq^{-\frac12}$ we find
$t_1=t_{N+1}=0$. Then (17a) and (17b) yield $X^+\psi_N=0$ and $X^-\psi_1=0$ so that $\psi_N$ and $\psi_1$ appear
as the highest and the lowest weight vectors respectively.

The important comment is in order: manipulating with quasi-momentum, the cyclic and the highest weight representations can
be transformed one into another. Such an interplay has been known for a long time \cite{arnaudon}.

\subsection{Functional representation}

Consider an $N$-component vector $f=(f_1,f_2,\ldots,f_N)$. Writing out the action of $X^\pm$, $K$, $K^{-1}$ on $f$ in
the component form we obtain
\babc
\begin{align}
(X^+f)_n&=e^{-i\bk\boa_1}q^{n-1}f_{n-1}+e^{-i\bk\boa_2}q^nf_{n+1},\\
\nn\\
(X^-f)_n&=e^{+i\bk\boa_1}q^{-n}f_{n+1}+e^{+i\bk\boa_2}q^{-n+1}f_{n-1},\\
\nn\\
(Kf)_n&=e^{+i\bk(\boa_1-\boa_2)}f_{n+2},\\
\nn\\
(K^{-1}f)_n&=e^{-i\bk(\boa_1-\boa_2)}f_{n-2},
\end{align}
\eabc
where the identification $f_{N+n}=f_n$ is assumed.

Introduce the interpolating function of the complex variable $f(z)$ such that
\be
f_n=f(e^{-\frac{i}{2}\sqrt3k_ya}q^{n-\frac12}).
\ee
Then the relations (20) can be written as
\babc
\begin{align}
X^+f(z)&=e^{-\frac{i}{2}k_xa}q^{-\frac12}zf(q^{-1}z)+\nn\\
\nn\\
&+e^{+\frac{i}{2}k_xa}q^{+\frac12}zf(qz),\\
\nn\\
X^-f(z)&=e^{+\frac{i}{2}k_xa}q^{-\frac12}z^{-1}f(qz)+\nn\\
\nn\\
&+e^{-\frac{i}{2}k_xa}q^{+\frac12}z^{-1}f(q^{-1}z),\\
\nn\\
Kf(z)&=e^{+ik_xa}f(q^2z),\\
\nn\\
K^{-1}f(z)&=e^{-ik_xa}f(q^{-2}z).
\end{align}
\eabc
These relations determine the functional representation of $U_q(sl_2)$.
Functional representation for the odd values of $\nu$ is constructed in Appendix B.

\section{The eigenvalue equation}

We study the eigenvalue equation
\be
\left\lgroup\ba{cc} 0 & \mathbb I+X^-(\bk)
\\\\
\mathbb I+X^+(\bk) & 0 \ea\right\rgroup
\left\lgroup\ba{c}\xi
\\\\
\zeta\ea\right\rgroup=E\left\lgroup\ba{c}\xi
\\\\
\zeta\ea\right\rgroup.
\ee

Employing the functional representation (22) the eigenvalue equation (23) can be rewritten in the form of two coupled equations
\babc
\begin{align}
\xi(z)&+e^{+\frac{i}{2}k_xa}q^{+\frac12}z\xi(qz)+\nn\\
\nn\\
&+e^{-\frac{i}{2}k_xa}q^{-\frac12}z\xi(q^{-1}z)=E\zeta(z),\\
\nn\\
\zeta(z)&+e^{+\frac{i}{2}k_xa}q^{-\frac12}z^{-1}\zeta(qz)+\nn\\
\nn\\
&+e^{-\frac{i}{2}k_xa}q^{+\frac12}z^{-1}\zeta(q^{-1}z)=E\xi(z).
\end{align}
\eabc
If any pair $(\xi,\zeta)$ is the eigenvector corresponding to the eigenvalue
$E=\lambda$, then the pair $(\xi,-\zeta)$ is the eigenvector corresponding
to the eigenvalue $E=-\lambda$.

The main disadvantage of (24) is the entanglement of $\xi(z)$ and $\zeta(z)$. This difficulty has been overcome in \cite{kohmoto}
by "squaring up" the equations (23): repeated application of $\mathscr H$ to (23) gives out $(\mathbb I+X^-)(\mathbb I+X^+)\xi=E^2\xi$
and $(\mathbb I+X^+)(\mathbb I+X^-)\zeta=E^2\zeta$ which in the functional representation appear as
\begin{widetext}
\babc
\begin{align}
&(3-E^2)\xi(z)+e^{+ik_xa}q\xi(q^2z)
+e^{+\frac{i}{2}k_xa}\big(q^{+\frac12}z+q^{-\frac12}z^{-1}\big)\xi(qz)+\nn\\
\nn\\
&+e^{-\frac{i}{2}k_xa}\big(q^{-\frac12}z+q^{+\frac12}z^{-1}\big)\xi(q^{-1}z)+e^{-ik_xa}q^{-1}\xi(q^{-2}z)=0,\\
\nn\\
&(3-E^2)\zeta(z)+e^{+ik_xa}q^{-1}\zeta(q^2z)
+e^{+\frac{i}{2}k_xa}\big(q^{+\frac12}z+q^{-\frac12}z^{-1}\big)\zeta(qz)+\nn\\
\nn\\
&+e^{-\frac{i}{2}k_xa}\big(q^{-\frac12}z+q^{+\frac12}z^{-1}\big)\zeta(q^{-1}z)+e^{-ik_xa}q\zeta(q^{-2}z)=0.
\end{align}
\eabc
\end{widetext}

Due to the "square up" trick some portion of information encoded in (24) is lost in (25). Here we propose essentially different approach allowing
to avoid this drawback (disadvantage) of the "square up" procedure. The key point of our consideration is the observation, that the system (24)
is invariant under the following transformation
\babc
\begin{align}
\xi(z)&\to(iz)^\omega\zeta(-z^{-1})\\
\nn\\
\zeta(z)&\to(iz)^\omega\xi(-z^{-1})
\end{align}
\eabc
where the parameter $\omega$ is set by $q^\omega e^{+ik_xa}=-1$. This symmetry allows to express the solutions to (24) as
\be
\left\lgroup\ba{c}\xi(z)\\\\\zeta(z)\ea\right\rgroup
=\left\lgroup\ba{c}f(z)\\\\ \pm(iz)^\omega f(-z^{-1})\ea\right\rgroup
\ee
where $f(z)$ satisfies the equation
\begin{align}
f(z)&+e^{+\frac{i}{2}k_xa}q^{+\frac12}zf(qz)+\nn\\
\nn\\
&+e^{-\frac{i}{2}k_xa}q^{-\frac12}zf(q^{-1}z)=\lambda(iz)^\omega f(-z^{-1}).
\end{align}
The signs "$\pm$" in (27) correspond to $E=\pm\lambda$, respectively.

We have thus reduced the system of equations (24) to a single equation (28) while $\xi(z)$ and $\zeta(z)$ are now expressed
via the unique function $f(z)$. In Appendix C we show that the system of equations (25) appears as a corollary of (28) meaning
that the later is the basic equation relevant to the problem under consideration.

We proceed to study the equation (28) at special values of momenta reducing the functional representation to the polynomial one.
Consider the monomials $f_j(z)=z^j$ forming the basis in the space of analytic functions. We then find
\babc
\begin{align}
X^+f_j(z)&=t_{j+1}f_{j+1}(z),\\
\nn\\
X^-f_j(z)&=t_jf_{j-1}(z),\\
\nn\\
Kf_j(z)&=q^{+2j}f_j(z),\\
\nn\\
K^{-1}f_j(z)&=q^{-2j}f_j(z),
\end{align}
\eabc
where $t_j$ has been defined by (18).

Setting $e^{+\frac{i}{2}k_xa}=iq^\frac12$ we find $X^-f_0=X^+f_{N-1}=0$ yielding that the subspace of $(N-1)$'th order polynomials
$f(z)=c_0+c_1z+\cdots+c_{N-1}z^{N-1}$ is the invariant subspace of the group $U_q(sl_2)$. Then the equations (24) appear as
\babc
\begin{align}
\xi(z)+iqz\xi(qz)-iq^{-1}z\xi(q^{-1}z)&=E\zeta(z),\\
\nn\\
\zeta(z)+iz^{-1}\zeta(qz)-iz^{-1}\zeta(q^{-1}z)&=E\xi(z),
\end{align}
\eabc
and admit the solutions in the form of $(N-1)$'th order polynomials. These polynomials are real, {\it i.e.} with real coefficients.
This statement can be verified by expanding $\xi(z)$ and $\zeta(z)$ in a series of $z^j$ and rewriting (30) for the expansion coefficients.

The invariance (26) holds for (30) provided $\omega=2J\equiv N-1$ while the equation (28) turns into
\be
f(z)+iqzf(qz)-iq^{-1}zf(q^{-1}z)=(-1)^J\lambda z^{2J}f(-z^{-1}),
\ee
generating $N$ eigenvalues $\lambda_1,\ldots,\lambda_N$. The factor of $(-1)^J$ can be absorbed in $\lambda$
and therefore we may drop it in what follows.

The equation (31) is derived for $\nu=even$. For $\nu=odd$ we come to the similar functional equation, but with the deformation
parameter $\tilde q=e^{\frac{i\pi(N-\nu)}{N}}$ instead of $q$ (see Appendix B). Therefore, provided $N=odd$ and $\nu=odd$
we have $N-\nu=even$, {\it i.e.} $\tilde q^N=1$. For that reason below we discuss only $\nu=even$.

Since $f(z)$ is a polynomial we can write down
\be
f(z)=\prod_{j=1}^{2J}(z-z_j),
\ee
where $z_1,z_2,\ldots,z_{2J}$ are the zeros of $f(z)$.

Substituting (32) into (31) and taking $z=-z_n^{-1}$ we obtain
\be
iz_n=\prod_{j=1}^{2J}\frac{1+qz_nz_j}{1+z_nz_j}-\prod_{j=1}^{2J}\frac{1+q^{-1}z_nz_j}{1+z_nz_j},
\ee
which is the honeycomb analog of the Bethe ansatz equation obtained in \cite{wiegmann} for square lattice. Due to the fact that (30)
is soluble in terms of real polynomials, the solutions to the equation (31) are also real polynomials. Correspondingly, if $z_n$ is a root
then $z^*_n$ must be the root as well. This comes in agreement with the invariance of (33) under the complex conjugation.

Rewriting (31) as
\be
\lambda=\prod_{j=1}^{2J}\frac{z-z_j}{1+zz_j}
+iqz\prod_{j=1}^{2J}\frac{qz-z_j}{1+zz_j}
-\frac{iz}{q}\prod_{j=1}^{2J}\frac{q^{-1}z-z_j}{1+zz_j},
\ee
and taking the limit $z\to0$ we find
\be
\lambda=\prod_{j=1}^{2J}z_j
-z\bigg[\sum_{j=1}^{2J}\bigg(\frac{1}{z_j}+z_j\bigg)-iq+\frac{i}{q}\bigg]
\prod_{j=1}^{2J}z_j+\mathscr O(z^2),
\ee
where the right hand side must be $z$-independent. Then
\be
\lambda=\prod_{j=1}^{2J}z_j
\ee
and
\be
\sum_{j=1}^{2J}\bigg(z_j+\frac{1}{z_j}\bigg)=iq+\frac{1}{iq}.
\ee

The analogous "sum rule" does exist in the case of the square lattice. Namely,
the functional relation derived in \cite{wiegmann} leads to
\be
\sum_{j=1}^{2J}\bigg(\mathcal Z_j+\frac{1}{\mathcal Z_j}\bigg)=0,
\ee
where $\mathcal Z_j$ are the zeros of the square lattice polynomials.

Taking now the limit $z\to\infty$ in (34) we find
\be
\lambda^2=1+i(q-q^{-1})\sum_{j=1}^{2J}z_j+\mathscr O\bigg(\frac{1}{z}\bigg),
\ee
where again the right hand side is $z$-independent yielding
\be
\lambda^2=1+i(q-q^{-1})\sum_{j=1}^{2J}z_j.
\ee

Using (37) we can rewrite (40) also as
\be
\lambda^2=3-q^2-q^{-2}-i(q-q^{-1})\sum_{j=1}^{2J}\frac{1}{z_j}.
\ee

We have thus derived the relation among an eigenvalue and the roots of the corresponding polynomial. The set of roots is determined by (33).
Once the roots of a polynomial are known, the corresponding eigenvalue can be calculated using any of (36), (40), (41).

Remark that the aforementioned invariance of (30) may be interpreted in terms of special conformal transformation
\be
z\to w=w(z)=-\frac{1}{z}.
\ee
Under this conformal map a quasi-primary field $\Phi(z)$ transforms as
\be
\Phi(z)\to\Phi'(w)=\bigg(\frac{dw}{dz}\bigg)^{-h}\Phi(z)
\ee
where $h$ is the corresponding conformal weight \cite{difrancesco}.

Taking $h=-J$ and expressing non-primed fields via the primed ones we find
\babc
\begin{eqnarray}
\Phi(z)=z^{2J}\Phi'(w),\\
\nn\\
\Phi(qz)=q^{-1}z^{2J}\Phi'(q^{-1}w),\\
\nn\\
\Phi(q^{-1}z)=qz^{2J}\Phi'(qw).
\end{eqnarray}
\eabc
Using (44) and rewriting the systems (30) in terms of primed fields we obtain
\babc
\begin{eqnarray}
\zeta'(w)+iqw\zeta'(qw)-iq^{-1}w\zeta'(q^{-1}w)=E\xi'(w),\\
\nn\\
\xi'(w)+iw^{-1}\xi'(qw)-iw^{-1}\xi'(q^{-1}w)=E\zeta'(w).
\end{eqnarray}
\eabc
Hence, we may identify $\zeta'(z)=\xi(z)$ and $\xi'(z)=\zeta(z)$,
{\it i.e.} $\zeta(z)=z^{2J}\xi(-z^{-1})$ and $\xi(z)=z^{2J}\zeta(-z^{-1})$.

\section{Numerical analysis}

We discuss the results of numerical analysis on how the roots are distributed over the complex plane. These studies are helpful for visualization
of the peculiarities hidden in the Bethe-like equation (33).

We first present the case of $\frac{\Phi}{\Phi_0}=\frac{2}{89}$ where we have $89$ eigenvalues and the corresponding $89$ polynomials
determined by (31). Each polynomial is of order of $N-1=88$, {\it i.e.} there are 88 roots for each eigenvalue. For demonstrative purposes
we present several characteristic cases, exhibiting how the root locations in the complex plane depend on eigenvalues. For the eigenvalue of
maximal magnitude the roots are arranged along two spirals related via the complex conjugation (figure 2a). For the eigenvalue of less
magnitude some of the roots (out of total 88) regroup into the new branch of circular form (figures 2b and 2c). For the eigenvalue of minimal
magnitude the spirals convert and the roots become arranged along two ovals (figure 2d).

\begin{widetext}

\begin{figure}
\includegraphics{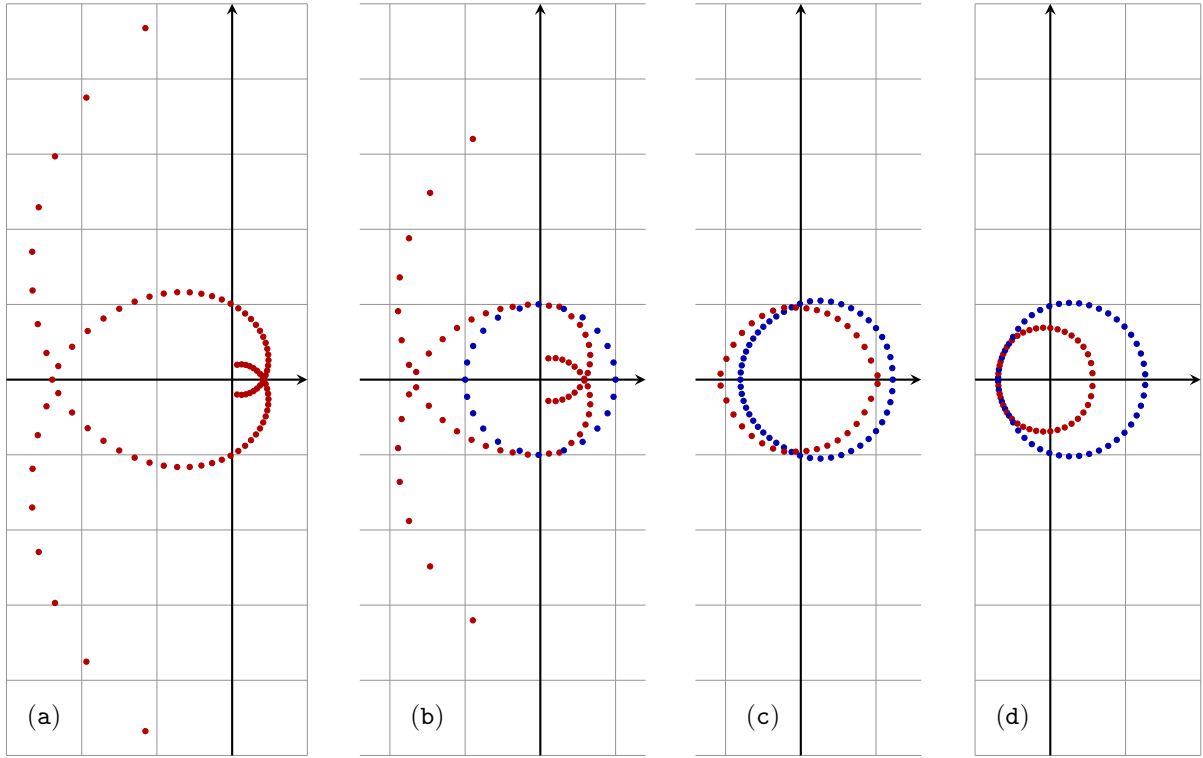}
\caption{Root distributions in the complex plane for $\frac{\Phi}{\Phi_0}=\frac{2}{89}$. There are $89$ eigenvalues and the $89$
corresponding polynomials, each of order of $N-1=88$. For each eigenvalue we have separate set comprising 88 roots of the corresponding
polynomial. The panel (a) exhibits the locations of the 88 roots for the eigenvalue of maximal magnitude. For the ones of less magnitude the
typical root distributions are shown on panels (b) and (c). The panel (d) corresponds to the eigenvalue of minimal magnitude where the spirals
are converted and the roots located along two ovals.}
\end{figure}

The case of $\frac{\Phi}{\Phi_0}=\frac{30}{89}\approx\frac13$ is presented in figure 3. In the case of maximal $\lambda^2$ the roots are
arranged in two groups (figure 3a): straight lines with equiangular separations of $\pi/3$ and the six points resembling hexagon vertices.
Few points are deviated from the straight lines, what can be explained by relatively small value of $N$. Passing to $\lambda^2$ of less magnitude
the roots regroup from the straight lines into a circle (figure 3b) which subsequently transforms into a hexagon (figure 3c) for the eigenvalue of
minimal magnitude. The points left beyond the hexagon can be treated as remnants of the straight lines.

We finally present the case of $\frac{\Phi}{\Phi_0}=\frac{22}{89}\approx\frac14$ in figure 4. Interplay between the two groups of roots is
again the case. One group consists of 8 straight lines with equiangular separation of $\pi/4$, and the other one resembling certain closed contour.
Changing $\lambda^2$ we observe that roots rearrange from one group into another and {\it vice versa}. For the eigenvalue of minimal magnitude
(figure 4c) the roots are located basically along the closed contour which appears to be an octagon.

Figure 3 ($\frac{\Phi}{\Phi_0}=\frac{30}{89}\approx\frac13$) and figure 4 ($\frac{\Phi}{\Phi_0}=\frac{22}{89}\approx\frac14$) demonstrate
that the root distributions are of high-accuracy symmetric with respect to the rotations by $\pi/3$ and $\pi/4$, respectively. These observations
provoke to think that in the limit $\nu\to\infty$, $N\to\infty$ with $\frac{\Phi}{\Phi_0}=\frac{\nu}{N}\to\frac13,\frac14$, the approximate symmetry
can become exact. In this light the limiting case of $\nu\to\infty$, $N\to\infty$ with $\frac{\nu}{N}\to finite$ is of special interest. We presume that
in this limit the equation (31) may comprise rich symmetry properties. We consider this issue as a matter of separate studies.

\begin{figure}
\includegraphics{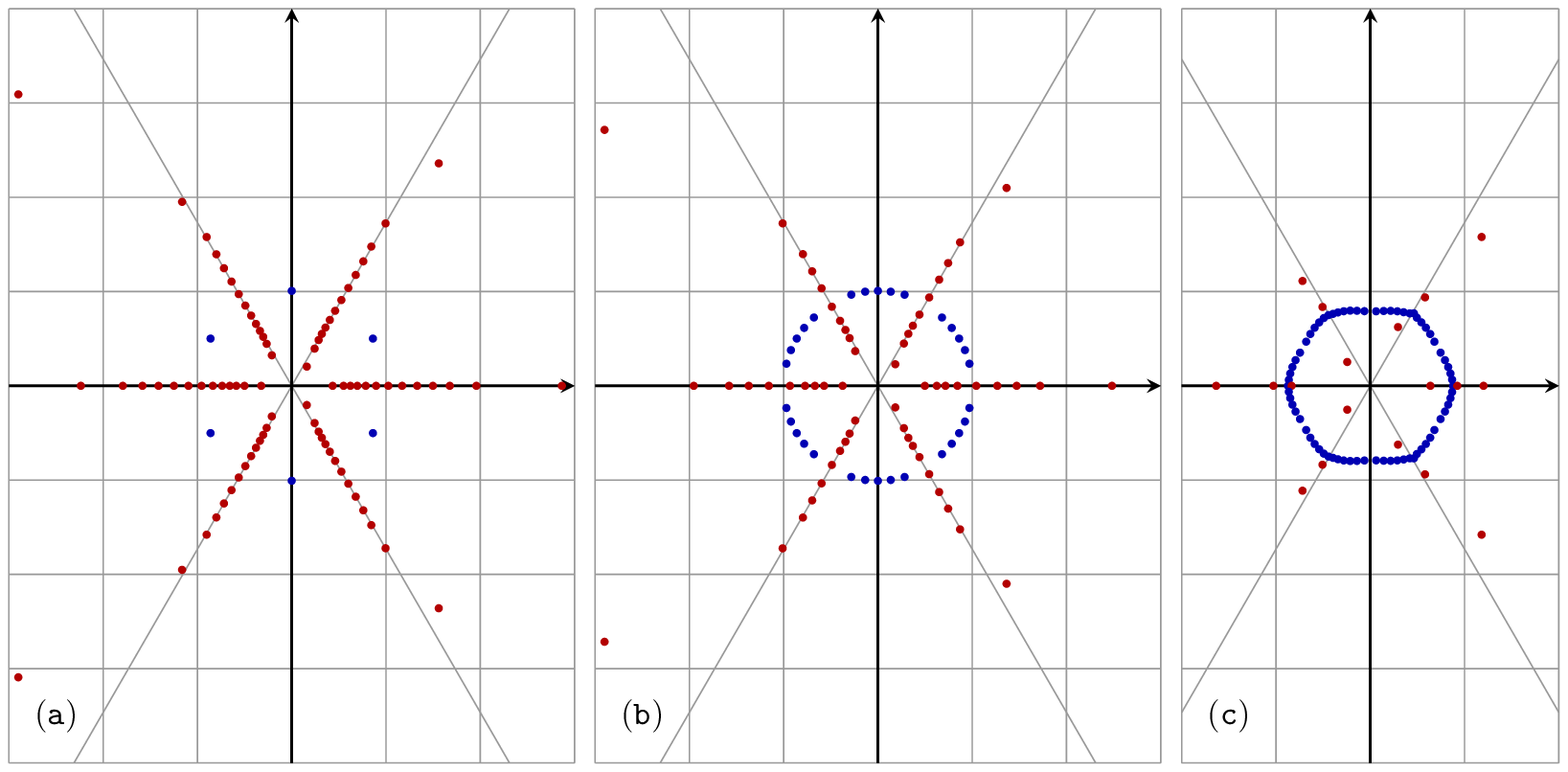}
\caption{Root distributions in the complex plane for $\frac{\Phi}{\Phi_0}=\frac{30}{89}\approx\frac13$. There are $89$ eigenvalues and the $89$
corresponding polynomials, each of order of $N-1=88$. For each eigenvalue we have separate set comprising 88 roots of the corresponding polynomial.
The panel (a) exhibits the locations of the 88 roots for the eigenvalue of maximal magnitude. The panel (c) corresponds to the eigenvalue of minimal magnitude.}
\end{figure}

\begin{figure}
\includegraphics{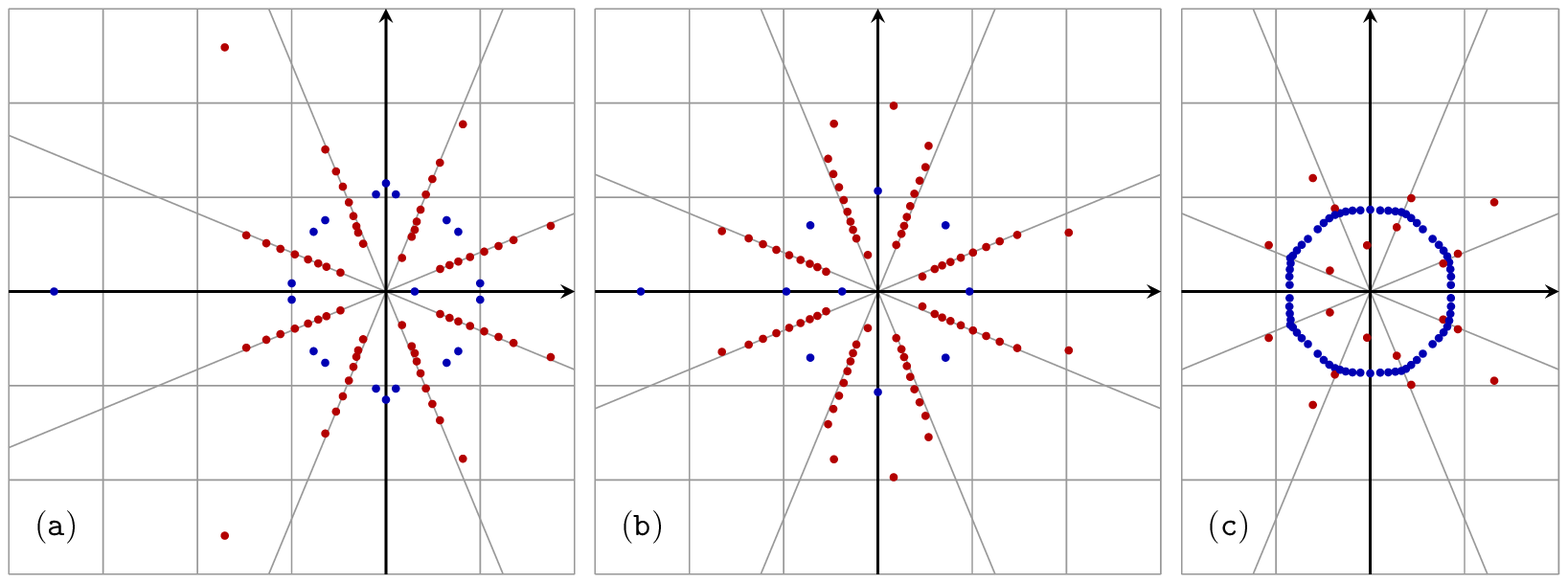}
\caption{Root distributions in the complex plane for $\frac{\Phi}{\Phi_0}=\frac{22}{89}\approx\frac14$. There are $89$ eigenvalues and the $89$
corresponding polynomials, each of order of $N-1=88$. For each eigenvalue we have separate set comprising 88 roots of the corresponding polynomial.
The panel (a) exhibits the locations of the 88 roots for the eigenvalue of maximal magnitude. The panel (c) corresponds to the eigenvalue of minimal magnitude.}
\end{figure}

\section{Conclusions}

We have considered the tight-binding model of spinless particles on a honeycomb lattice in magnetic field. The corresponding one-particle Hamiltonian turns out
to be expressible in terms of the generators of the quantum group $U_q(sl_2)$. We have shown that varying the momentum $\bk$ the cyclic representation of
$U_q(sl_2)$ can be continuously deformed into the highest weight one. Using the functional representation of $U_q(sl_2)$ the Harper equation is reformulated
as a system of two entangled functional equations. Tracing out certain symmetry of the entangled system we reduce the two equations to a single equation (31)
admitting solutions in terms of real polynomials. Bethe-like equations determining the roots of polynomials are obtained and the roots are solved out numerically
for the different eigenvalues $\lambda$ and different magnetic fluxes $\Phi$. Using graphical plots we have shown that the root distributions are highly organized
and arranged in regular geometric figures.

We have stated the quantum group $U_q(sl_2)$ in terms of $q$-deformed commutation relations (12) which are compatible with $q$-deformed co-product
structure providing tensor product representations. The representations discussed in Sections 3 and 4 correspond to one-particle states, while the higher
representations constructed via the aforementioned $q$-deformed product structures will be helpful for describing many-body quantum states.

The procedure described in Section 4 enables to explicitly construct the complete set of orthogonal one-particle wave functions $|\Psi_{\bk,j}\rangle$
($j=1,\ldots,N$) at specific values of momenta set by $e^{+\frac{i}{2}k_xa}=iq^\frac12$, where the functional representation of the quantum group becomes
reduced to polynomial representation. As pointed out in subsection 3.1 the similar reduction takes place at other values of $k_x$ given by
$e^{+\frac{i}{2}k_xa}=\pm iq^{\frac12-j_0}$ where $j_0=0,1,\ldots,N-1$. Consequently for each aforementioned value of $k_x$ we are able to construct
the set of one-particle wave functions which can be subsequently employed for calculating variety of matrix elements
$\langle\Psi_{\bk_1,j_1}|\cdots|\Psi_{\bk_2,j_2}\rangle$. In particular, calculations of the matrix elements of the Coulomb operator are in progress.

\section{Acknowledgments}

We would like to thank A. Elashvili and M. Jibladze for helpful discussions. One of the authors (M.E.) is grateful to P. Sorba for valuable communications.
The research was supported in part by the Georgian NSF through the Grants ST09/09-280 (G.I.J) and ST08/4-405 (M.E., G.Ts.).

\appendix

\section{}

Consider first the case of $\nu=even$. We then have
\be
|\bk_0|=\frac{\frac12\nu}{N}\frac{4\pi}{\sqrt3a},
\ee
where $\frac12\nu$ is integer, {\it i.e.} $|\bk_0|$ is the multiple of the $N$'th part of the Brillouin zone. We then split the integration area in
(4) into $N$ horizontal strips of the height $\frac{1}{N}\frac{4\pi}{\sqrt3a}$. Introducing $N$-component vectors ($\mu=A,B$)
\be
\Psi_\mu(\bk)=\{c_\mu(\bk),c_\mu(\bk-\bk_0),\ldots,c_\mu(\bk-N\bk_0+\bk_0)\},
\ee
and combining $\Psi_A$ and $\Psi_B$ into $\Psi=\{\Psi_A,\Psi_B\}$
we eventually come to (7).

In the case of $\nu=odd$ we have to write (A1) as
\be
|\bk_0|=\frac{\nu}{2N}\frac{4\pi}{\sqrt3a},
\ee
{\it i.e.} $|\bk_0|$ is now the multiple of the $(2N)$'th part of the Brillouin zone. Correspondingly we split the integration area of
(4) into $2N$ equal horizontal strips and rewrite (4) as
\be
H=\int_{{\tt FBZ}/(2N)}\Big\{\mathbb H^\dag(\bk)+\mathbb H(\bk)\Big\}\dk,
\ee
where ${\tt FBZ}/(2N)$ denotes any separate strip and
\begin{align}
\mathbb H
&=\sum_{n=1}^{2N}
c^\dag_A(\bk-n\bk_0+\bk_0)c^{}_B(\bk-n\bk_0+\bk_0)+\nn\\
\nn\\
&+\sum_{n=1}^{2N}\xi^{}_n(\bk)
c^\dag_A(\bk-n\bk_0+\bk_0)c^{}_B(\bk-n\bk_0)+\nn\\
\nn\\
&+\sum_{n=1}^{2N}\zeta^{}_n(\bk)
c^\dag_A(\bk-n\bk_0)c^{}_B(\bk-n\bk_0+\bk_0),
\end{align}
with $\xi_n(\bk)=q^{-n}e^{+i\bk\boa_1}$ and $\zeta_n(\bk)=q^{-n}e^{+i\bk\boa_2}$.

We then decouple the summations into $1\leqslant n\leqslant N$ and $N+1\leqslant n\leqslant2N$ and introduce ($\mu=A,B$)
\be
\chi_{\mu n}(\bk)=\frac{1}{\sqrt2}c_\mu(\bk-n\bk_0+\bk_0)+\frac{(-1)^ni^N}{\sqrt2}c_\mu(\bk-N\bk_0-n\bk_0+\bk_0).
\ee
Constructing
$\Psi_\mu=\{\chi^{}_{\mu 1},\chi^{}_{\mu 2},\ldots,\chi^{}_{\mu N}\}$
and $\Psi=\{\Psi_A,\Psi_B\}$ we come to (7).

\section{}

Discussing the case of $\nu=even$ in subsection 3.2. we wrote down (20) reflecting the action of the group generators on a vector
$f=(f_1,f_2,\ldots,f_N)$. The same relations are valid also for $\nu=odd$ only if the aperiodic conditions $f_{n+N}=(-1)^{n+1}f_n$
are assumed. Provided these conditions are accepted, the functional representation cannot be constructed as $f_n=f(q^n)$ any longer.

In order to overcome this difficulty we introduce the basis where $K$ and $K^{-1}$ are diagonal. For $\nu=odd$ these basis states are given by
\be
u_j\equiv\frac{1}{\sqrt{N}}\big(
\alpha_j,
-\alpha_j^{J+2},
\alpha_j^2,
-\alpha_j^{J+3},
\alpha_j^3,
\ldots,
\alpha_j^{J},
-\alpha_j^{2J+1},
\alpha_j^{J+1}
\big)^T,
\ee
where $\alpha_j\equiv q^{2j+1}$ and $N=2J+1$.

The actions of group generators on $u_j$ read as
\begin{align}
X^+u_j&=\kappa
\big(e^{-i\bk\boa_1}\tilde q^{j+1}+e^{-i\bk\boa_2}\tilde q^{-j-1}\big)u_{j+1},\\
\nn\\
X^-u_j&=\kappa^{-1}
\big(e^{+i\bk\boa_1}\tilde q^{-j}+e^{+i\bk\boa_2}\tilde q^{j}\big)u_{j-1},\\
\nn\\
Ku_j&=-e^{+ik_xa}\tilde q^{-2j-1}u_j,\\
\nn\\
K^{-1}u_j&=-e^{-ik_xa}\tilde q^{2j+1}u_j,
\end{align}
where $\tilde q=-q^*=e^{i\pi(1-\nu/N)}$ and $\kappa=(-\tilde q)^{1-J}$.

Remark that due to $N=odd$ we have $\tilde q^N=1$, {\it i.e.} the obstacle related to $q^N=-1$ is removed by reformulating the action of $X^\pm$
in terms of $\tilde q$. Correspondingly, the relations (B.2) -- (B.5) require the periodic conditions $u_{j+N}=u_j$ which on the other hand are really
met by (B.1). Physically this is related to $\Phi\to\Phi_0-\Phi$.

We now pass to another basis $v_j$ set by
\be
u_j=N^{-\frac12}\sum_{n=1}^N \tilde q^{jn}v_n.
\ee
Then the relations (B.2) and (B.3) turn into
\begin{align}
\kappa^{-1}X^+v_j
&=e^{-i\bk\boa_1}\tilde q^{j}v_{j-1}+e^{-i\bk\boa_2}\tilde q^{j}v_{j+1},\\
\nn\\
\kappa X^-v_j
&=e^{+i\bk\boa_1}\tilde q^{-j-1}v_{j+1}+e^{+i\bk\boa_2}\tilde q^{1-j}v_{j-1},
\end{align}
which are the analogues of (20a) and (20b).

Every $N$-component column can be expanded as
\be
f\equiv\left(\ba{c} f_1\\\vdots\\f_N\ea\right)
=\phi_1v_1+\phi_2v_2+\cdots+\phi_Nv_N,
\ee
so the action of group generators on $f$ is now understood
as acting on the columns $v_j$.

Alternatively, the action of group operators on $f$ can be understood as modifying the corresponding coordinates $\phi=(\phi_1,\ldots,\phi_N)$.
Provided $X^\pm$ act on $v_j$ as set by (B.7) and (B.8), the equivalent actions on $\phi$ are given by
\begin{align}
\kappa^{-1}(X^+\phi)_j&=e^{-i\bk\boa_1}\tilde q^{j+1}\phi_{j+1}+e^{-i\bk\boa_2}\tilde q^{j-1}\phi_{j-1},\\
\nn\\
\kappa(X^-\phi)_j&=e^{+i\bk\boa_1}\tilde q^{-j}\phi_{j-1}+e^{+i\bk\boa_2}\tilde q^{-j}\phi_{j+1}.
\end{align}
Introducing here
\be
\phi_n=f(\kappa e^{-\frac{i}{2}\sqrt3k_ya}\tilde q^n),
\ee
we come to the functional representation
\begin{align}
X^+f(z)&=e^{-\frac{i}{2}k_xa}\tilde q z f(\tilde qz)+e^{+\frac{i}{2}k_xa}\tilde q^{-1}z f(\tilde q^{-1}z),\\
\nn\\
X^-f(z)&=e^{+\frac{i}{2}k_xa}z^{-1}f(\tilde q^{-1}z)+e^{-\frac{i}{2}k_xa}z^{-1}f(\tilde qz).
\end{align}
Setting $e^{-\frac{i}{2}k_xa}=i$, the eigenvalue equation appears as
\begin{align}
\xi(z)+i\tilde qz\xi(\tilde qz)-i\tilde q^{-1}z\xi(\tilde q^{-1}z)&=E\zeta(z),\\
\nn\\
\zeta(z)+iz^{-1}\zeta(\tilde qz)-iz^{-1}\zeta(\tilde q^{-1}z)&=E\xi(z),
\end{align}
and coincides with (30).

\section{}

Performing the replacement $z\to-z^{-1}$ in (28) we obtain
\be
f(-z^{-1})-e^{+\frac{i}{2}k_xa}q^{+\frac12}z^{-1}f(-qz^{-1})
-e^{-\frac{i}{2}k_xa}q^{-\frac12}z^{-1}f(q^{-1}z^{-1})
=\lambda(iz)^{-\omega}f(z).
\ee
Multiplying both sides by $\lambda(iz)^{\omega}$ we pass to
\be
\lambda(iz)^{\omega}f(-z^{-1})-e^{+\frac{i}{2}k_xa}q^{+\frac12}z^{-1}\lambda(iz)^{\omega}f(-qz^{-1})
-e^{-\frac{i}{2}k_xa}q^{-\frac12}z^{-1}\lambda(iz)^{\omega}f(q^{-1}z^{-1})=\lambda^2f(z).
\ee
where the first terms of left hand side can be replaced using (28). The second and third terms can be replaced in accord with
\begin{align}
\lambda(iz)^\omega f(-qz^{-1})
&=q^\omega f(q^{-1}z)+q^\omega e^{+\frac{i}{2}k_xa}q^{-\frac12}zf(z)
+q^\omega e^{-\frac{i}{2}k_xa}q^{-\frac32}zf(q^{-2}z)\\
\nn\\
\lambda(iz)^\omega f(-q^{-1}z^{-1})
&=q^{-\omega}f(qz)+q^{-\omega}e^{+\frac{i}{2}k_xa}q^{+\frac32}zf(q^2z)
+q^{-\omega}e^{-\frac{i}{2}k_xa}q^{+\frac12}zf(z)
\end{align}
which can be obtained from (28) by $z\to qz$ and $z\to q^{-1}z$, respectively.

Performing these manipulations in (C.1) and taking into account $q^\omega e^{+ik_xa}=-1$ we obtain
\begin{eqnarray}
&(3-\lambda^2 )f(z)+e^{+ik_xa}qf(q^2z)
+e^{+\frac{i}{2}k_xa}\big(q^{+\frac12}z+q^{-\frac12}z^{-1}\big)f(qz)+\nn\\
\nn\\
&+e^{-\frac{i}{2}k_xa}\big(q^{-\frac12}z+q^{+\frac12}z^{-1}\big)f(q^{-1}z)
+e^{-ik_xa}q^{-1}f(q^{-2}z)=0.
\end{eqnarray}
which coincides with (25a). Reformulating for $F(z)\equiv(iz)^\omega f(-z^{-1})$ we come to (25b).

\newpage

\end{widetext}


\begin{thebibliography}{15}
\bibitem{azbel} M.Ya. Azbel, {\it Zh. Eksp. Teor. Fiz.} {\bf 46} (1964) 929; {\it Soviet Physics JETF} {\bf 19} (1964) 636.
\bibitem{hofstadter} D.R. Hofstadter, {\it Phys. Rev.} {\bf B14} (1976) 2239.
\bibitem{claro79} F.H. Claro and G.H. Wannier, {\it Phys. Rev.} {\bf B19} (1979) 6068.
\bibitem{claro81} F. Claro, {\it Phys. Status Solidi} {\bf B104} (1981) K31.
\bibitem{hasegawa} Y. Hasegawa, P. Lederer, T.M. Rice and P.B. Wiegmann, {\it Phys. Rev. Lett.} {\bf 63} (1989) 907.
\bibitem{semenoff} G.W. Semenoff, {\it Phys. Rev. Lett.} {\bf 53} (1984) 2449.
\bibitem{rammal} R. Rammal, {\it J. Physique} {\bf 46} (1985) 1345.
\bibitem{wiegmann} P.B. Wiegmann and A.V. Zabrodin, {\it Phys. Rev. Lett.} {\bf 72} (1994) 1890.
\bibitem{faddeev} L.D. Faddeev and R.M. Kashaev, {\it Commun. Math. Phys.} {\bf 169} (1995) 181.
\bibitem{hatsugai} Y. Hatsugai, M. Kohmoto and Y.-Sh. Wu, {\it Phys. Rev.} {\bf B53} (1996) 9697.
\bibitem{kohmoto} M. Kohmoto and A. Sedrakyan, {\it Phys. Rev.} {\bf B73} (2006) 235118.
\bibitem{geim} A.K. Geim and K.S. Novoselov, {\it Nature Materials} {\bf 6} (2007) 183.
\bibitem{kassel} C. Kassel, {\it Quantum Groups}, Graduate Texts in Mathematics vol. 155 (Springer-Verlag 1995).
\bibitem{arnaudon} P. Roche and D. Arnaudon, {\it Lett. Math. Phys.} {\bf 17} (1989) 295.
\bibitem{difrancesco} Ph. Di Francesco, P. Mathieu and D. S\'{e}n\'{e}chal, {\it Conformal Field Theory} (Springer-Verlag 1997).
\end{thebibliography}
\end{document}